\documentstyle[prl,aps,multicol]{revtex}
\newenvironment{Fig}{
\begin{figure}[h]
\noindent\begin{minipage}[t]{3.4in}
\begin{center}
\leavevmode
\epsfxsize=3 true in
}
{
\end{center}
\end{minipage}
\end{figure}
}
\begin{document}
\bibliographystyle{prsty}
\input epsf
\title{Time Dependent Development of the Coulomb Gap}
\author{Clare C. Yu}
\address{
Department of Physics and Astronomy, University of California,
Irvine, Irvine, California 92697}
\date{\today}
\maketitle
\begin{abstract}
We show that the time development of the Coulomb gap in a Coulomb glass
can involve very long relaxation times due to electron rearrangement 
and hopping. We find that an applied magnetic field reduces the
rate of electron hopping and, hence, Coulomb gap formation. These
results are consistent with recent conductance experiments on
thin semiconducting and metallic films. 

\end{abstract}


\begin{multicols}{2}
\narrowtext
The competition between interactions and disorder result in
glassy dynamics that are often associated with very long relaxation times
extending over many decades. 
One might not
expect the same to be true in an electronic system 
since electrons typically respond very quickly. 
However, in this paper we show that
in the presence of strong disorder, electrons
can indeed have very long relaxation times. 
This occurs in a Coulomb
glass which is an insulator with randomly placed electrons
that have Coulomb interactions. Heavily doped semiconductors
and disordered metals are examples of such systems. 
Coulomb interactions between localized
electrons result in a so--called Coulomb gap in the single
particle density of states that is centered at the Fermi energy
\cite{Pollak70,Efros75,efrosbook}.
We have done a calculation in which we follow the time development of
the Coulomb gap.
In order to produce this gap, electron rearrangement must occur
and the associated hopping can involve very long time scales.

These long relaxation times are consistent with recent experiments
on thin semiconducting \cite {Ovadyahu97,Vaknin98a}  
and metallic \cite{Martinez-Arizala98} films which have shown
that in the presence of strong disorder, electronic systems
can relax very slowly. These films were grown on insulating substrates
which separated them from a gate electrode that regulated the electron
density, and hence the chemical potential, of the film. 
The conductance $G$ was measured as a function
of the gate voltage $V_{G}$. If $V_{G}$ sat at a particular value, $V_{o}$,
for a long time and then was varied over a range of voltages,
there was a dip in the conductance centered at $V_{o}$ \cite{comment:notSC}.
We identify this dip with the Coulomb gap in the density of states
because the value of the conductance depends on the density of states at
the Fermi energy \cite{Vaknin98a,goldman}. In Mott's picture 
of variable range hopping,
the hopping conductivity increases when the density of states
at the Fermi energy increases, since there are then more states
to which an electron at the Fermi energy can hop \cite{efrosbook}. 
We identify sweeping $V_{G}$ with varying the chemical potential without 
allowing time for equilibration. In effect the sweeps scan the
density of states. Thus we expect the conductance to increase with the
density of states and hence as the gate voltage
$V_{G}$ moves away from $V_{o}$. 

In the experiments,
if the gate voltage was changed suddenly from, say, $V_{o}$
to $V_{1}$, the conductance had a very fast
initial rise, followed by a period of rapid relaxation, which in turn
was followed by a long period of very slow relaxation. In some cases
the relaxation was logarithmic in time. Our interpretation of
this is that when the gate voltage is changed, the Fermi energy
changes, and time dependent relaxations arise because the system must dig
a new hole in the density of states at the new Fermi energy and
remove the old hole at the old Fermi energy. Indeed,
Vaknin, Ovadyahu and Pollak \cite{Vaknin98a} found that subsequent
sweeps of the gate voltage revealed that the old dip in the conductance
at $V_{o}$ fades with time while a new dip centered at $V_{1}$ increased
with time. 
The dip in the conductance and the long time relaxation
were present only at very low temperatures, not at 
higher temperatures ($T\stackrel{>}{\sim} 20$ K). 

Ovadyahu and Pollak \cite{Ovadyahu97} also found that in a magnetic 
field ($H=9$ Tesla) the long time relaxation rate associated with
a change in the gate voltage was even slower. 
In addition to the spin mechanism that they propose \cite{Ovadyahu97},
this result is consistent with the fact that the magnetic field reduces the
spatial extent of the electron wavefunctions in the
directions transverse to the field. This reduces the wavefunction
overlap of neighboring electrons, resulting in a decrease of the
electron hopping rate and hence a decrease in the rate at which
a Coulomb gap forms. This is confirmed by our calculations.

Our model of the Coulomb glass follows that of Baranovski\u{i},
Shklovski\u{i}, and A.L. \'{E}fros (BSE) \cite{Baranovskii80}. 
In this model, the electrons occupy the sites of a periodic lattice,
and the number of electrons is half the number of sites.
Each site has a random onsite energy $\phi_{i}$ chosen from a uniform
distribution extending from $-A$ to $A$. Thus, $g_{o}$, the
density of states without interactions, is flat. A site can contain 0 or 1
electron. In order to follow the time development of the Coulomb
gap, we assume that the Coulomb interactions are turned on at time
$t=0$. The Hamiltonian can be written as
\begin{equation}
H = \sum_{i}\phi_{i}n_{i} + \sum_{i>j}\frac{e^{2}}{\kappa r_{ij}}n_{i}n_{j}
\theta(t)
\label{eq:hamiltonian}
\end{equation}
where the occupation number $n_{i}$ equals 
$\frac{1}{2}$ if site $i$ is occupied
and $-\frac{1}{2}$ if site $i$ is unoccupied, $e$ is the electron charge,
$\kappa$ is the dielectric constant,
and the step function $\theta(t)$ is 0 for $t<0$ and 1 for $t\geq 0$. 

The Coulomb gap arises because the
stability of the ground state with respect to single
electron hopping from an occupied site $i$ to an unoccupied
site $j$ requires \cite{efrosbook}
\begin{equation}
\Delta_{i}^{j} = \varepsilon_{j}-\varepsilon_{i}-\frac{e^{2}}{\kappa r_{ij}}>0
\label{eq:stability}
\end{equation}
where the single--site energy $\varepsilon_{i}=\phi_{i}+\sum_{j}
\frac{e^{2}}{\kappa r_{ij}}n_{j}$. 
So we need to subtract from the density of states
those states which violate this stability condition.
This leads to \cite{Baranovskii80,Burin95}
\begin{eqnarray}
g(\varepsilon,& t)&= g_{o}  \prod_{j>i} \left( 1-a_{o}^{3}\int_{-A}^{A}
d\varepsilon^{\prime} g(\varepsilon^{\prime},t)
\theta(\frac{e^{2}}{\kappa r_{ij}}+\varepsilon-
\varepsilon^{\prime})\right.      \nonumber   \\
  &   & \left. \phantom{\int_{-A}^{A}}
  F(n_{i}^{\prime}=1,n_{j}^{\prime}=0)\theta (t-\tau_{ij}
  (\varepsilon^{\prime},\varepsilon, r_{ij}))\right)
\label{eq:dosprod}
\end{eqnarray}
where the single--site energy $\varepsilon_{i}=\varepsilon$, 
$\varepsilon_{j}=\varepsilon^{\prime}$, and
$a_{o}$ is the lattice constant. $n_{i}^{\prime}=n_{i}+1/2$; so
$n_{i}^{\prime}=1$ if site i is occupied and $0$ if site $i$ is unoccupied.
$F(n_{i}^{\prime},n_{j}^{\prime})$ 
is the probability
that donors $i$ and $j$ have occupation numbers $n_{i}^{\prime}$ and
$n_{j}^{\prime}$, respectively, while all other sites have their ground state
occupation numbers $\tilde{n}_{k}^{\prime}$. $\tau_{ij}^{-1}$ is the number
of electrons which jump from site $i$ to site $j$ per unit time.
$\theta(t-\tau_{ij})$ represents the fact that at time $t$, the
primary contributions to the change in the
density of states will be from those hops for which $\tau_{ij}<t$
\cite{comment:exp}. In writing eq. (\ref{eq:dosprod}), we
assume that these hops together with phonons
have equilibrated the system as much as is possible at time $t$.
The hopping rate $\tau_{ij}^{-1}$ is given by \cite{efrosbook}
\begin{equation}
\tau_{ij}^{-1}=\gamma_{ij}^{o}\exp(-\frac{2r_{ij}}{a})
[1+N(\Delta^{j}_{i})] F(n_{i}^{\prime}=1,n_{j}^{\prime}=0)
\label{eq:tau}
\end{equation}
where 
$a=\kappa a_{B}$ is the effective Bohr radius of a
donor, and $a_{B}$ is
the usual Bohr radius ($a_{B}=\hbar^2/me^2$). We will set
the mass $m$ equal to the electron mass so that $a_{B}=0.529\AA$.
$N(\Delta^{j}_{i})$ is the phonon occupation factor and reflects
the contribution of phonon assisted hopping.  
We are also allowing for spontaneous
emission of phonons since we are considering a nonequilibrium situation
in which electrons hop in order to lower their energy. 
The coefficient $\gamma_{ij}^{o}$ is given by \cite{efrosbook}
\begin{equation}
\gamma_{ij}^{o}=\frac{E_{1}^{2}|\Delta^{j}_{i}|}{\pi d s^{5}\hbar^{4}}
\left[\frac{2e^{2}}{3\kappa a}\right]^{2}\frac{r_{ij}^{2}}{a^2}
\left[1+\left(\frac{\Delta^{j}_{i}a}{2\hbar s}\right)^{2}\right]^{-4}
\label{eq:gamma0}
\end{equation}
where $E_{1}$ is the deformation potential, $s$ is the speed of sound,
and $d$ is the mass density. 
Following BSE, we can derive a self--consistent equation for the density
of states $g(\varepsilon,t)$: 
\begin{eqnarray}
& &g(\varepsilon,t) =  g_{o}\exp \left\{-\frac{1}{2}\int_{-A}^{A}
d\varepsilon^{\prime}g(\varepsilon^{\prime},t) 
 \int_{a_{o}}^{\infty}dr 4\pi r^{2} \right. \nonumber       \\ 
& &\left. F(n(\varepsilon)=1,n(\varepsilon^{\prime})=0)
\theta(\frac{e^{2}}{\kappa r}+\varepsilon-\varepsilon^{\prime})
\theta(t-\tau(\varepsilon^{\prime},\varepsilon,r)) \right\} 
\label{eq:dos}
\end{eqnarray}
At low energies large distances play an important role and so we
have replaced the sum by an integral over $r$ in the exponent.
The origin is at site $i$.
$n(\varepsilon)$ is the occupation probability of a site
with energy $\varepsilon$. $\tau(\varepsilon^{\prime},\varepsilon,r)$
is given by (\ref{eq:tau}) with $r_{ij}$ replaced by $r$, 
$\varepsilon_{i}$ replaced by $\varepsilon$, and $\varepsilon_{j}$ replaced
by $\varepsilon^{\prime}$. 

Since it is not clear how the stability condition of eq. (\ref{eq:stability})
can be applied to finite temperatures, we will confine our calculations to
the case of $T=0$. In this case
the electron occupation factor $F(n_{i}^{\prime}=1,n_{j}^{\prime}=0)=1$ and 
the phonon occupation factor $N(\Delta^{j}_{i})=0$. We can
solve eq. (\ref{eq:dos}) iteratively on the computer. 
For the first iteration we start with $g(\varepsilon^{\prime},t)=g_{o}$,
and calculate $g(\varepsilon,t)$. This is then used as the input
for $g(\varepsilon^{\prime},t)$ in the next iteration.
Because successive iterations converge by alternating above and below the
correct answer with decreasing amplitude, 
after the first two iterations we use the average of the
input and output of a given iteration as the input for the next iteration.
After 11 iterations the typical difference between successive iterations
is typically less than 1 part in $10^{5}$. We set the Fermi energy
$\mu=0$.
Because there is particle--hole symmetry, we only need to 
calculate $g(\varepsilon,t)$ for $\varepsilon < 0$.
Figure \ref{fig:dosT0}a shows the density of states $g(\varepsilon,t)$
as a function of energy at different times, while Figure \ref{fig:dosT0}b
shows $g(\varepsilon,t)$ as a function of time at different energies.
Notice the development of the Coulomb gap occurs over many decades
in time. The functional form of the time dependence of
$g(\varepsilon,t)$ varies with the energy $\varepsilon$ and
with $g_o$. For example, at the Fermi energy $g(\mu,t)\sim \ln t$ for
$g_{o}=2\times 10^{5}$ states/K--\AA$^{3}$ and $g(\mu,t)\sim t^{-0.05}$ for
$g_{o}=6.25\times 10^{5}$ states/K--\AA$^{3}$.
After an infinite amount of time, the density of states at the
Fermi energy $\mu$
goes to zero and $g(\varepsilon)\sim\varepsilon^2$.
For finite times, $g(\varepsilon)\sim|\varepsilon|$ in the vicinity of
the Fermi energy, though there will be thermal smearing at finite temperatures.
For different values of the initial density of states $g_o$,
we find the same qualitative behavior as a function of
time with the depth of the dip $g_o-g(\mu,t)$ 
increasing as $g_o$ increases. 
Figure \ref{fig:width} shows that the width $W$ of the dip 
increases with $g_o$. 
Experimentally the width $W$ increases with the carrier concentration $n$ 
\cite{Vaknin98a}. This is consistent with our results
since the noninteracting density of states $g_o$ increases with
$n$, though other parameters such as $\kappa$ may also depend on $n$. 
The range of widths in Figure \ref{fig:width} 
is comparable to that deduced from experiment \cite{Vaknin98a}. 

The temporal development of
the Coulomb gap is qualitatively consistent with the experimental
observation of the long time relaxation of the conductance after the
gate voltage $V_{G}$ has been changed. 
The exact relation between the conductance and the density of
states is difficult to acertain in this case because the system is not in
equilibrium. However, it is reasonable to assume that 
the conductance reflects the density of states at the Fermi energy.
A well known example is Mott's formula for conductivity $\sigma$ due
to variable range hopping \cite{efrosbook}.
\begin{equation}
\sigma=\sigma_{o}\exp\left[-\left(\frac{T_{o}}{T}\right)^{\frac{1}{4}}\right]
\label{eq:mott}
\end{equation}
where $T_{o}=\alpha/(k_{B}g(\mu)a^{3})$, $\alpha$ is a numerical
constant, and $g(\mu)$ is the density of states at the Fermi energy.
While strictly speaking this equilibrium formula does not apply
to our nonequilibrium situation, we see qualitatively that 
an increase (decrease) in $g(\mu)$ leads to an increase (decrease)
in the conductivity. In the experiments, rapidly sweeping the gate voltage
$V_{G}$ varies the chemical potential without allowing time for
equilibration. Relating the conductance to the density of states means
that the sweeps over $V_{G}$ scan the density of states. To get a 
qualitative feel for this connection, we will use
eqn. (\ref{eq:mott}). We identify $g(\mu)$ with $g(\varepsilon)$
and use $T_{o}=\alpha/(k_{B}g(\varepsilon)a^{3})$.
For most of the scan the density of states has the linear form
$g(\varepsilon)=g(\varepsilon_o)+\alpha (\varepsilon-\varepsilon_o)$
where $\varepsilon_o$ and the slope $\alpha$ are constants.
The exponent of 1/4 is appropriate for this case.
The experiments on indium oxide
\cite{Ovadyahu97} were done at 4.11 K, so we set $T=4$ K and use the
$g(\varepsilon)$ shown in Figure \ref{fig:dosT0}. The result is shown
in Figure \ref{fig:condT0}.

Ovadyahu and Pollak noticed that when
a magnetic field is applied to their indium oxide 
films, the relaxation rate decreases and
the magnetoresistance is positive \cite{Ovadyahu97}. They attributed
this to a reduced hopping rate resulting
from the fact that a polarized spin cannot hop onto
a site that is already occupied. Our scenario suggest an additional
mechanism since the magnetic field reduces the wavefunction overlap
in the direction transverse to the field. 
The reduced overlap means a lower hopping rate and a longer relaxation time.
Shklovskii and Efros \cite{Shklovskii82,Shklovskii83a,Shklovskii83b}
studied the effect of a magnetic field on variable range hopping. Since the
average hopping distance far exceeds the mean distance $R$ between 
impurities, the hopping electron scatters from many other donor
sites. As a result, when the magnetic field is transverse
to the direction of tunneling, the wavefunction decays as
$\exp(-\rho/b)$. Here we have adopted cylindrical coordinates with
the magnetic field along the $z$ axis and $\rho$ is the radial coordinate
transverse to the $z$ axis. The parameter $b=\lambda/|\ln A|$ where
the magnetic length
$\lambda=\sqrt{c\hbar/e H}$, and $A$ describes the scattering
and depends on the magnetic field $H$. For all values of $H$,
Shklovski\u{i} \cite{Shklovskii83b} has shown that
$b$, and hence the wavefunction overlap, decreases as the 
magnetic field increases. The functional form of $A$ depends on
the strength of the field. In the indium oxide experiments 
\cite{Ovadyahu97} 
$H=9$ Tesla. Since this is in a weak field regime where $\lambda \gg a$ 
and $R\ll \lambda^{2}/a$, 
we can make the approximation
$b\approx a[1-(a/\lambda)^{4/3}]$ where $a$ is the effective
Bohr radius. For a weak field we expect the hopping rate
to go as $\tau_{ij}^{-1}\sim\gamma_{ij}^{o}\exp[-2r_{ij}/f(z/\rho_{ij})]$
where the function $f(0)\equiv b$ and $f(\infty)\equiv a$.
It is difficult to make a quantitative
comparison to the density of states at $H=0$ since we do not know the
prefactor $\gamma_{ij}^{o}$. To get a qualitative feel for the
effect of the magnetic field, we can use the $H=0$ form of $\gamma_{ij}^{o}$
found in eqn. (\ref{eq:gamma0}), and in the equation (\ref{eq:tau})
for $\tau_{ij}^{-1}$, we replace $\exp(-2r_{ij}/a)$ with $\exp(-2r_{ij}/b)$.
This is a reasonably good approximation since 
for a field of 9 Tesla and a dielectric constant of 10,
$a=5.29$ \AA  and $b=5.162$ \AA.
The $T=0$ result for the density of states at
$H=0$ and $H=9$ Tesla is shown in the inset
of Figure \ref{fig:width}. Both curves start at the
same value of $g(\varepsilon=\mu)=g_o$ at time $t=0$. 
Notice that the curve
at 9 Tesla is slightly above the zero field curve indicating that
the relaxation is slower in a magnetic field. This is qualitatively
consistent with experiment \cite{Ovadyahu97}. To differentiate between
our mechanism and spin effects, note that our mechanism predicts that the
magnetoresistance is greater when the field is perpendicular to the current
than when it is parallel, whereas the magnetoresistance should be
isotropic in the field if spin effects dominate.

The experiments found that the dip in the conductance as a function
of gate voltage $V_{G}$ and the long relaxation times of the
conductance following a change in $V_{G}$ were present only at low
temperature. These features were not observed at $T\stackrel{>}{\sim}$
20 K. From eqs. (\ref{eq:dosprod}) and (\ref{eq:tau}), we see that
an increase in temperature will affect $g(\varepsilon,t)$ in two ways. 
First the
thermal smearing of the occupation factor $F(n_{i}^{\prime},n_{j}^{\prime})$
will fill in the Coulomb gap to some extent. Secondly, as the number
of phonons increases with temperature, there is an increase in the
phonon assisted hopping of electrons. We expect that this leads to 
a rapid rearrangement of electrons on time scales that are too short
to observe experimentally. As a result, no dips in the conductance
and no long time relaxation were seen experimentally at higher temperatures.
It is difficult to calculate these effects because it is not clear
how to generalize the stability condition (\ref{eq:stability})
to finite temperatures. In addition the
system is not in equilibrium and hence temperature is not well defined
for the electrons. However the absence of the conductance dips
at higher temperatures is consistent with our scenario.

To summarize, we have shown that the time development of the Coulomb gap
in a Coulomb glass can involve very long time scales due to electron
hopping and rearrangement. These results are
consistent with conductance experiments on disordered semiconducting
and metallic films. Although we have only 
considered single electron hops, these hops are dependent upon
previous hops of other electrons through their cumulative effect on
the single particle density of states. We expect multielectron 
processes to also contribute to the conductance, particularly at long
time scales.

I would like to thank Herv\'e Carruzzo for helpful discussions.
This work was supported in part by ONR grant
N00014-96-1-0905 and by funds provided by the University of
California for the conduct of discretionary research by Los
Alamos National Laboratory.

\begin{Fig}
\epsfbox{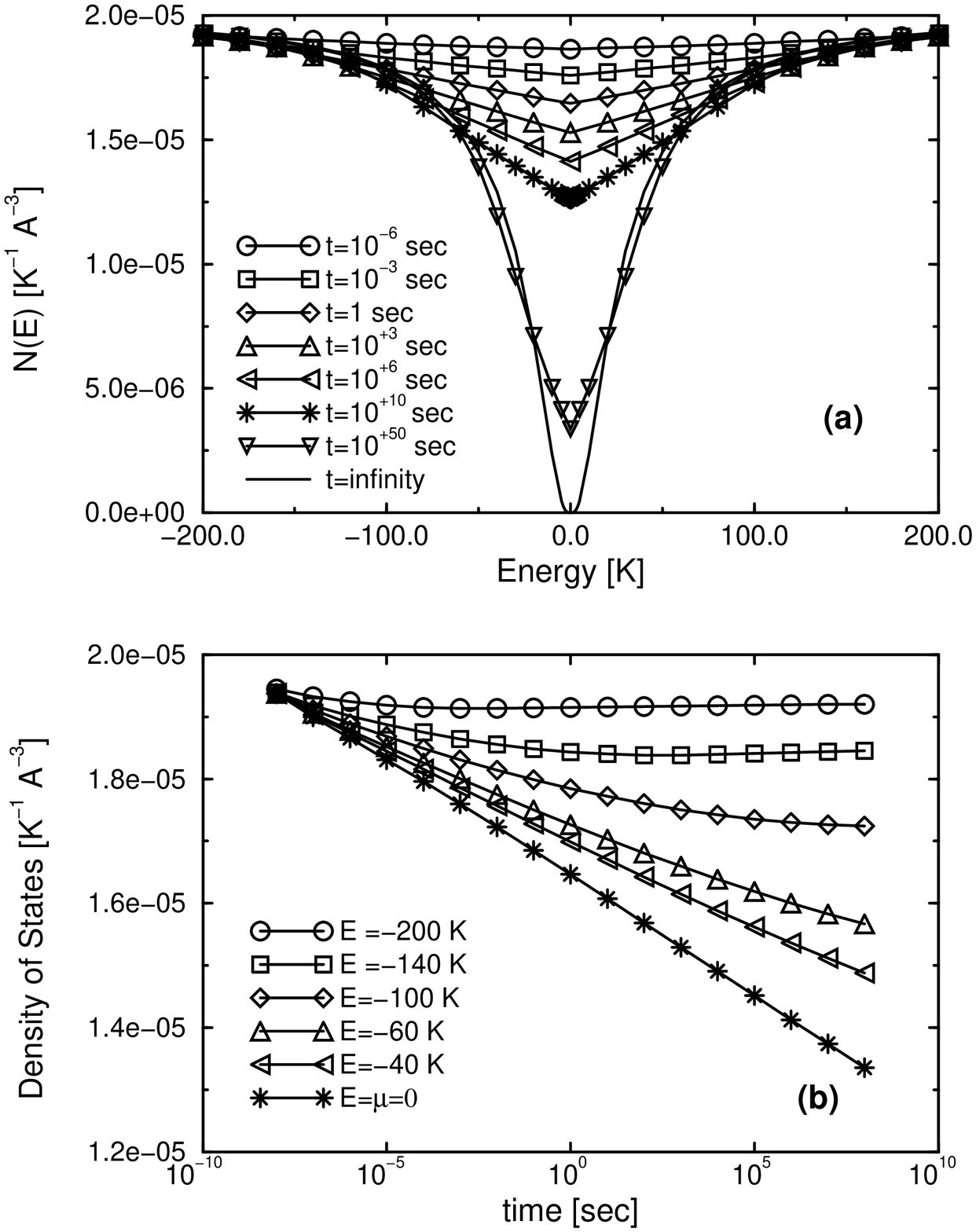}
\caption{(a) Density of states $g(\varepsilon)$ as a function of energy
for different times. (b) Density of states as a function of time for
various energies. Parameters used are $g_{o}=2\times 10^{5}$ 
states/K--\AA$^{3}$,
$T=0$, $A=10^4$ K, $\kappa=10$,
$d=7.18$ g/cm$^{3}$, $s=5.0\times 10^{5}$ cm/sec, $E_{1}=5 \times 10^{3}$ K,
and $a_{o}=4$ \AA. The density $d$ is chosen to be that of In$_{2}$O$_{3}$.
The energy is measured from the Fermi energy $\mu=0$.
}
\label{fig:dosT0} 
\end{Fig}
\begin{Fig}
\epsfbox{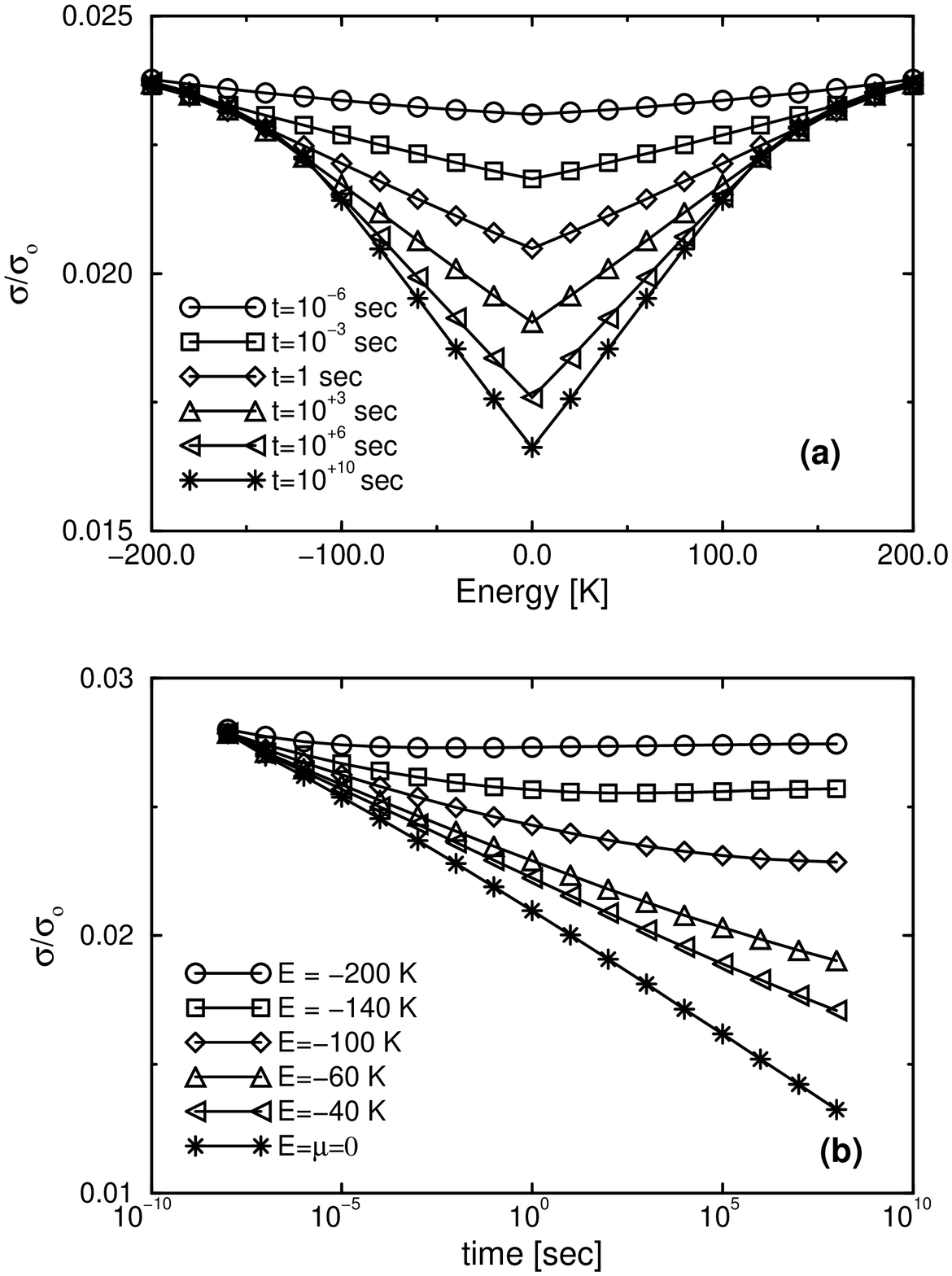}
\caption{(a) Dimensionless conductivity $\sigma/\sigma_o$ 
as a function of energy
for different times. (b) Dimensionless conductivity $\sigma/\sigma_o$ 
as a function of time for
various energies. The conductivity of both (a) and (b) 
are calculated using Mott's formula (\ref{eq:mott}) with
$T=4$ K, $a=5.29177 \AA$ and $\alpha=2.23$. 
The rest of the parameters are the same as in Figure \protect\ref{fig:dosT0}.
}
\label{fig:condT0}
\end{Fig}

\begin{Fig}
\epsfbox{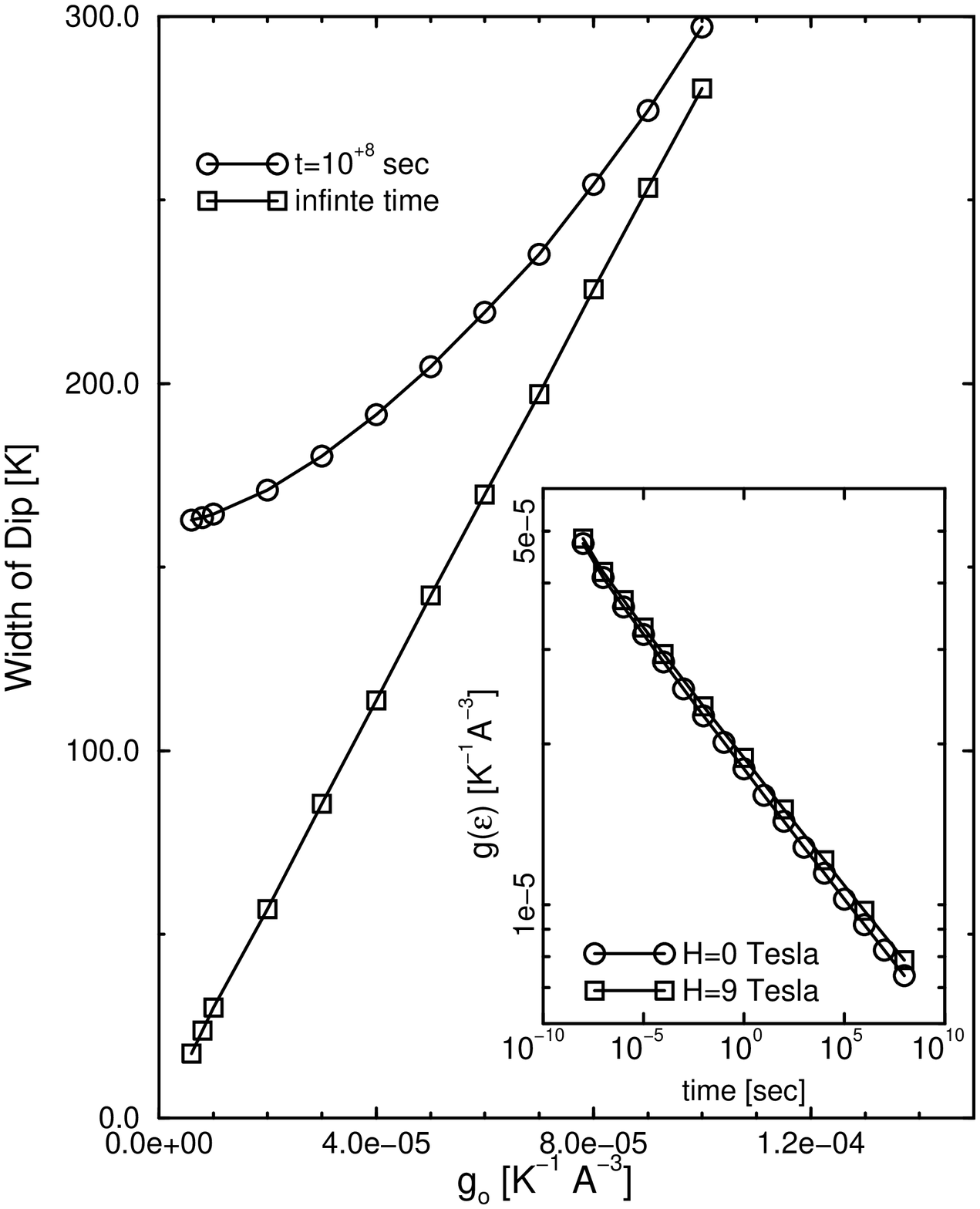}
\caption{Width of the Coulomb gap as a function of $g_o$ at
$t=10^{+8}$ sec and at infinite time. The width
is measured at halfway between the minimum $g(\mu)$ and the maximum $g_o$
in the density of states. 
Inset: Density of states $g(\varepsilon)$ as a function of time
for different values of the magnetic field $H$. For both curves the
energy $\varepsilon=\mu$ and $g_o=6.25\times 10^{-5}$ 
states/K--\AA$^{3}$. 
The rest of the parameters are the same as in Figure \protect\ref{fig:dosT0}
for both the inset and the main figure.}
\label{fig:width}
\end{Fig}

\end{multicols}

\begin{thebibliography}{10}
\bibitem{comment:notSC}
In this paper we will be concerned with highly resistive films which are
  insulating, not superconducting, at T=0.
\bibitem{goldman}
A similar scenario was qualitatively discussed in
  \protect\cite{Martinez-Arizala98}.
\bibitem{comment:exp}
The exact form of the cutoff is not important; for example, replacing
  $\theta(t-\tau_{ij})$ with $[1-\exp(-t/\tau_{ij})]$ affects our results
  negligibly.
\end{thebibliography}
\end{document}